\documentclass[useAMS,usedcolumn,usenatbib,usegraphicx]{mn2e}

\usepackage{epsfig,graphicx,natbib}
\usepackage{./reference/mycite}
\usepackage{amssymb}
\usepackage{gensymb}
\usepackage{amsfonts}
\usepackage{amsmath}
\usepackage{color}
\usepackage{lscape,longtable}
 
\begin{document}

\title[Reddened Quasars from VHS and \textit{WISE}]{Hyper-luminous Reddened Broad-Line Quasars at $z\sim$2 from the VISTA Hemisphere Survey and \textit{WISE} All Sky Survey} 
\author[M. Banerji et al.]{ \parbox{\textwidth}
{Manda Banerji$^{1,2}$\thanks{E-mail: mbanerji@ast.cam.ac.uk}, Richard G. McMahon$^{1,3}$, Paul C. Hewett$^{1}$, Eduardo Gonzalez-Solares$^{1}$, Sergey E. Koposov$^{1,4}$   
}
  \vspace*{6pt} \\
$^{1}$Institute of Astronomy, University of Cambridge, Madingley Road, Cambridge, CB3 0HA, UK.\\
$^{2}$Department of Physics \& Astronomy, University College London, Gower Street, London WC1E 6BT, UK. \\
$^{3}$Kavli Institute for Cosmology, University of Cambridge, Madingley Road, Cambridge, CB3 0HA, UK.\\
$^{4}$Moscow MV Lomonosov State University, Sternberg Astronomical Institute, Moscow 119992, Russia. \\
} 

\maketitle

\begin{abstract} 

We present the first sample of spectroscopically confirmed heavily reddened broad-line quasars selected using the new near infra-red VISTA Hemisphere Survey and \textit{WISE} All-Sky Survey. Observations of four candidates with $(J-K)>2.5$ and $K\le16.5$ over $\sim$180 deg$^2$, leads to confirmation that two are highly dust-reddened broad-line Type 1 quasars at z$\sim$2. The typical dust extinctions are A$_V\sim$2--2.5 mags. We measure black-hole masses of $\sim$10$^{9}$M$_\odot$ and extinction corrected bolometric luminosities of $\sim$10$^{47}$ erg/s, making these among the brightest Type 1 quasars currently known. Despite this, these quasars lie well below the detection limits of wide-field optical surveys like the SDSS with $i_{AB}>22$. We also present \textit{WISE} photometry at 3--22$\mu$m, for our full sample of spectroscopically confirmed reddened quasars including those selected from the UKIDSS Large Area Survey (Banerji et al. 2012a). We demonstrate that the rest-frame infrared SEDs of these reddened quasars are similar to UV-luminous Type 1 quasars with significant hot dust emission and starburst quasar hosts like Mrk231. The average 12$\mu$m flux density of our reddened quasars is similar to that of the recently discovered HyLIRG \textit{WISE}1814+3412 ($z=2.452$) at similar redshifts, with two of our reddened quasars also having comparable 22$\mu$m flux densities to this extreme HyLIRG. These optically faint, heavily reddened broad-line quasars are therefore among the most mid infrared luminous galaxies at $z\sim2$, now being discovered using \textit{WISE}.       

\end{abstract}


\section{INTRODUCTION}

Dust-reddened broad-line quasars could represent the crucial missing link between massive ultraluminous starbursts identified in far infrared and submillimeter surveys, and UV-luminous quasars in optical surveys. The study of these rare transitioning systems, has gained new impetus with the advent of sensitive wide-field infrared surveys such as the UKIDSS Large Area Survey (LAS) \citep{Lawrence:07}, VISTA Hemisphere Survey (VHS; McMahon et al. in prep) and Wide Infrared Survey Explorer (\textit{WISE}) All Sky Survey \citep{Wright:10}. In our previous work (\citet{Banerji:12}; B12 hereafter), we used the UKIDSS LAS to find a population of heavily dust-reddened ($(J-K)>2.5$ (Vega)) quasars at $z\sim$2 corresponding to the main epoch of black-hole growth in the Universe. This new sample of 12 confirmed reddened quasars was the first to be assembled using a pure near infrared (NIR) selection and pushed 2 mags fainter than previous studies using 2MASS, which had selected quasars that were considerably less red ($(J-K)>1.7-2$) \citep{Glikman:07, Cutri:01}

Sensitive NIR surveys at 1--2$\mu$m are well suited to searching for broad-line quasars at $z\sim$2, with significant amounts of dust in their host galaxy.  At these redshifts, the NIR colours sample a portion of the rest-frame spectral energy distribution (SED) at $\sim$0.3--0.8$\mu$m that is very sensitive to dust extinction within the quasar host, but not to the hot-dust emission from the molecular torus at 1--3$\mu$m \citep{Hyland:82, Neugebauer:87}. Ground-based NIR imaging also allows discrimination of luminous Type 1 broad-line quasars that typically appear unresolved, and the less luminous Type 2 sources where the host galaxy is often resolved. 

%The new VISTA telescope in Chile is carrying out the most sensitive NIR all-sky survey ever undertaken. The new VISTA Hemisphere Survey (VHS) represents a tenfold increase in sensitivity compared to the UKIDSS-LAS and a hundredfold increase in sensitivity compared to 2MASS over $\sim$5000 deg$^2$ of high-latitude sky. This data therefore offers an unprecedented opportunity to conduct a detailed study of the reddened broad-line quasar population. Recently, the all-sky release from the new \textit{WISE} satellite has provided another very large dataset within which to search for luminous AGN populations \citep{Assef:10, Stern:12}. Mid infra-red selection techniques have been employed for some time in order to find populations of obscured quasars that are absent from optical surveys \citep{Lacy:04, Stern:05}. These selection methods however do not discriminate between Type 1 and Type 2 AGN. 

Our aim in this work is to use initial data from the new VISTA Hemisphere Survey and the \textit{WISE} All Sky Survey completed this year, to devise a colour selection criterion for reddened broad-line quasars and subsequently test this selection method via spectroscopic follow-up of a small pilot sample. The infra-red colour selection scheme we use, should in principle be effective at isolating dust-reddened broad line quasars and the current study offers the first practical demonstration of this selection using new data from VHS and \textit{WISE}. In addition, we also investigate the rest-frame infra-red SEDs of our full sample of reddened ($(J-K)>2.5$) quasars using data from the \textit{WISE} All Sky Release. Throughout this paper we assume a flat concordance cosmology with H$_0$=70 kms$^{-1}$ Mpc$^{-1}$, $\Omega_M=0.3$ and $\Omega_\Lambda=0.7$. All magnitudes are in the Vega system unless otherwise stated. 

%The AB-Vega conversions are $Y$=0.618, $J$=0.937, $H$=1.384, $K$=1.839, $W1$=2.683, $W2$=3.319, $W3$=5.242, $W4$=6.604.

%All magnitudes are in the Vega system unless otherwise stated. The AB-Vega conversions are $Y$=0.618, $J$=0.937, $H$=1.384, $K$=1.839, $W1$=2.683, $W2$=3.319, $W3$=5.242, $W4$=6.604.     

\section{PHOTOMETRIC DATA}

\label{sec:phot}

\subsection{VISTA Hemisphere Survey (VHS)}

VHS is a NIR photometric survey being conducted using the VISTA telescope in Chile, that aims to survey 18,000 deg$^2$ of the southern celestial hemisphere to a depth 30 times fainter than the 2MASS survey in at least two wavebands ($J$ and $K$). In the South Galactic Cap, 4500 deg$^2$ will be imaged deeper, including the $H$-band, and will have supplemental deep $grizY$ imaging provided by the Dark Energy Survey. The remainder of the high galactic latitude sky will be imaged at $YJHK$ and combined with optical photometry from the VST-ATLAS survey. In this work, we use $\sim$180 deg$^2$ data from the VHS VST-ATLAS overlap region at 205$^{\degree}<$RA$<245^{\degree}$ and $-9^{\degree}<$DEC$<-4.5^{\degree}$. The VHS data in this region reaches average 5$\sigma$ $K$-band depths of 18.2 (Vega).  

%One of the main science goals of the VHS high-latitude surveys is the discovery of populations of dust obscured quasars as well as quasars at the highest redshifts (z$>6$) corresponding to the epoch of reionization (e.g. \citet{Mortlock:11}). These quasars \textit{drop out} in typical wide area flux-limited optical surveys such as the SDSS and are identified on the basis of their optical-NIR colours. 

%Automated searches for these rare populations of quasars are, however, plagued by the number of spurious sources and stellar contaminants that dominate wide-field survey data. These contaminants need to be well understood so that reliable selection criteria can be defined to isolate a manageable number of quasar candidates for spectroscopic follow-up. A secondary goal of our study is therefore also to better understand the new infra-red data from the VHS so as to inform future searches for both high-redshift and dust-obscured quasars. In this work, we use the VHS-ATLAS data for our reddened quasar selection over a small region of $\sim$180deg$^2$ centred on RA=15hr and DEC=2.25.

\subsection{WISE}

The \textit{WISE} satellite has conducted an all-sky survey at mid infrared wavelengths between 3.4$\mu$m and 22$\mu$m \citep{Wright:10}. \textit{WISE} specific mid-IR colour selection criteria for isolating both populations of obscured and unobscured AGN have been devised \citep{Assef:10, Stern:12, Assef:12}. These authors propose a colour-cut of [W1$-$W2]$\gtrsim$0.8--0.85 with a more conservative cut of [W1$-$W2]$\geq$0.7 suffering from less incompleteness but also prone to higher levels of contamination from populations of dusty and normal star-forming galaxies. These colour criteria have been shown to be extremely effective in selecting AGN at intermediate redshifts but become incomplete when H$\alpha$ moves into the W1 band at $z\gtrsim3.4$. The \textit{WISE} selection identifies AGN independent of their reddening and so needs to be combined with additional colour information at shorter wavelengths in order to distinguish highly reddened quasars from those that are unobscured at UV and optical wavelengths. In addition, morphological information from ground-based NIR surveys can be used to separate dusty broad-line quasars from Type 2 AGN. 

\subsection{This Sample}

Our aim is to test the effectiveness of a joint VHS+\textit{WISE} colour selection in isolating highly reddened quasars at z$\sim$2. We
therefore identify reddened quasar candidates that satisfy the
following selection criteria:

\begin{itemize}

\item{$(J-K)>2.5$ and 13$<K<17$ (Vega). A bright sample with $13<K\le$16.5 (Vega) is prioritized for spectroscopic follow-up. The $(J-K)$ colour-cut corresponds roughly to $E(B-V)>0.5$ at $z\sim2$.}

\item{Unresolved or stellar-like point source classifications in the VHS NIR images. This corresponds to \textit{mergedclass}=$-$1 in the VHS catalogues.}

\item{Signal-to-noise of $>$10 in the $K$-band in order to ensure reliable morphological classifications.}

\item{Candidates are required to be red in all the NIR bands - i.e. $J>H$ \& $H>K$.}

\end{itemize}

The above NIR only selection criteria produces $\sim$600 candidates with $K<17$, 447 of which are at $K\le16.5$. Many of these sources are expected to be spurious. In order to clean the sample, it is matched to the \textit{WISE} All-Sky source catalogue using a matching radius of 3$^{\prime \prime}$ and requiring a $>$5$\sigma$ detection in both the \textit{WISE} 3.4 and 4.6$\mu$m bands. This reduces the candidate list to 321 sources. The VHS+\textit{WISE} detected sources are then required to satisfy the additional criterion of [W1$-$W2]$\geq$0.85, which produces 11 candidate reddened quasars with $13<K<17$, seven of which are at $13<K\le16.5$. Finally, the images for all candidates are visually inspected to create the sample for spectroscopic follow-up. This visual inspection eliminates two sources with a close neighbour in the $J$-band, which may be affecting the $J$-band photometry, leaving five candidates, which are summarised in Table \ref{tab:sample}. 

\begin{figure*}
\begin{center}
\includegraphics[scale=0.5,angle=0]{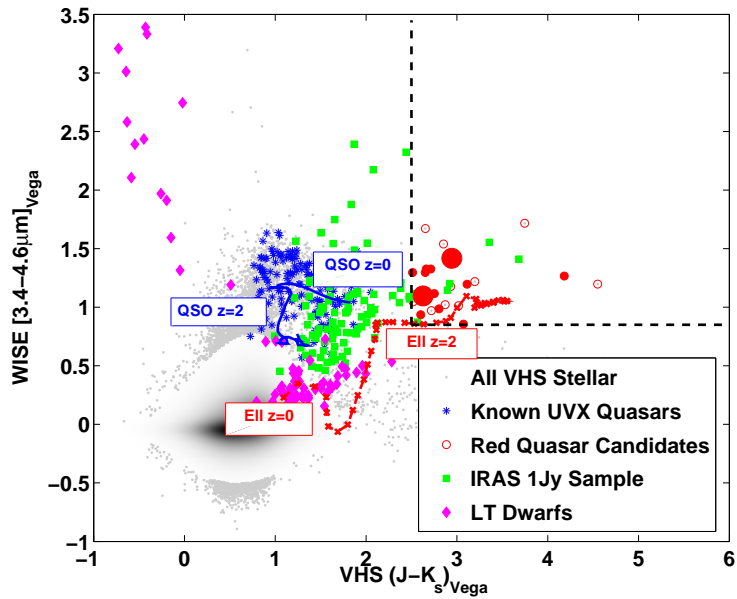}
\caption{$(J-K)$ versus (W1-W2) colour selection of our red quasar candidates. All stellar objects over 180 deg$^2$, are shown in grey. The greyscale represents the density of these objects while the 1\% of outliers in the distribution are shown as the individual grey points. Known UV-luminous quasars, local Ultraluminous Infrared Galaxies (ULIRGs) from the IRAS 1Jy sample and known LT dwarf stars have also been plotted. We also show the tracks of a typical unreddened quasar as well as an elliptical galaxy template with a formation redshift of $z=5$. Our reddened quasar candidates are marked as the red circles. The small filled circles represent the spectroscopically confirmed sample from B12, the two large filled circles are the 2 new confirmed quasars presented in this work and the open circles represent all candidates presented in this study down to $K<$17.}
\label{fig:select}
\end{center}
\end{figure*}

Before presenting results of our spectroscopic follow-up, we first look at the expected effectiveness of our colour selection criteria presented above. In Figure \ref{fig:select} we show our red quasar candidates from VHS+\textit{WISE}, with $13<K<$17 in the $(J-K)$ versus (W1-W2) colour-colour plane as well as our sample of already confirmed such reddened quasars from B12. Also shown are all point-like sources over the same area, down to the same flux limit in the $K$-band, 142 known UV-luminous quasars with $K<$17 from the SDSS Data Release 7 \citep{Schneider:10}, which have been matched to VHS and \textit{WISE}, 114 local ultraluminous infrared galaxies from the IRAS 1Jy sample \citep{Veilleux:02} and a sample of 73 L and T dwarf stars with UKIDSS-LAS and \textit{WISE} photometry \citep{Gelino:04}. We also plot the tracks of a UV-luminous quasar, and an evolving elliptical galaxy template modelled as a simple stellar population with $z_{\rm{form}}=5$, using an updated version of \citet{BC:03} and the publicly available web-based tool EzGAL\footnote{http://www.baryons.org/ezgal/}.

The reddened quasars are significantly redder in terms of their $(J-K)$ colours compared to the bulk of point sources in the VHS as well as the cool L and T dwarf stars. The reddened quasars are also considerably redder than known UVX quasars in the SDSS. Luminous infrared galaxies could enter our selection box as illustrated by the comparison to the IRAS 1Jy sample in Figure \ref{fig:select} but our morphological classification requirement of point sources in the $K$-band means that these ULIRGs would have to be extremely compact and/or at high redshifts. Finally, compact elliptical galaxies too could enter our selection box but only at $z>2$ where they are unlikely to exist in large numbers and extremely unlikely to be brighter than $K<17$.  

\begin{table}
\begin{center}
\caption{Sample of $13<K\le16.5$ and $(J-K)>2.5$ reddened quasar candidates from VHS and \textit{WISE}.}
\label{tab:sample}
\begin{tabular}{lcccc}
\hline \hline
Name & RA & DEC & K$_{\rm{Vega}}$ & Redshift \\
\hline
\hline
%VHSATLJ1345-0829 & 13:45:17.90 & $-$08:29:57.5 & 12.64 & 0.47 \\
VHSJ1350$-$0503 & 13:50:37.24 & $-$05:03:59.4 & 15.97 & 2.176 \\
VHSJ1409$-$0830 & 14:09:29.08 & $-$08:30:58.7 & 16.50 & 2.300 \\
VHSJ1504$-$0442 & 15:04:46.25 & $-$04:42:41.4 & 16.50 & No Spec \\
VHSJ1518$-$0501 & 15:18:37.85 & $-$05:01:38.3 & 16.13 & -- \\
VHSJ1543$-$0439 & 15:43:21.91 & $-$04:39:08.6 & 16.43 & -- \\
\hline
\end{tabular}
\end{center}
\end{table}

\section{SPECTROSCOPIC FOLLOW-UP}

Spectra were taken for four out of our five reddened quasar candidates using GNIRS, the NIR spectrograph on Gemini-North. We used GNIRS in cross-dispersed mode in order to obtain the widest wavelength coverage. Each target was observed for 20 minutes using 4 exposures of 5 minutes with the target dithered along the slit in a classical ABBA pattern for purposes of sky subtraction. We used the short camera (0.15''/pix) and the 31.7 lines/mm grating with a 1$^{\prime\prime}$ slit. This provided coverage over 0.9--2.5$\mu$m across six orders. Only the two reddest orders are used here corresponding to the atmospheric windows at 1.6$\mu$m and 2.2$\mu$m. The instrumental resolution is measured to be $\sim$10\AA\@ or 160 km/s from the sky lines. All data were reduced using standard GNIRS pipelines (v1.11) and utilising the Image Reduction and Analysis Facility (IRAF)
%\footnote{IRAF is distributed by the National Optical Astronomical Observatories, which are operated by the Association of Universities for Research in Astronomy Inc., under cooperative agreement with the National Science Foundation}.  

\begin{figure*}
\begin{center}
\centering
\begin{minipage}[c]{1.00\textwidth}
\begin{tabular}{cc}
\includegraphics[scale=0.5,angle=0]{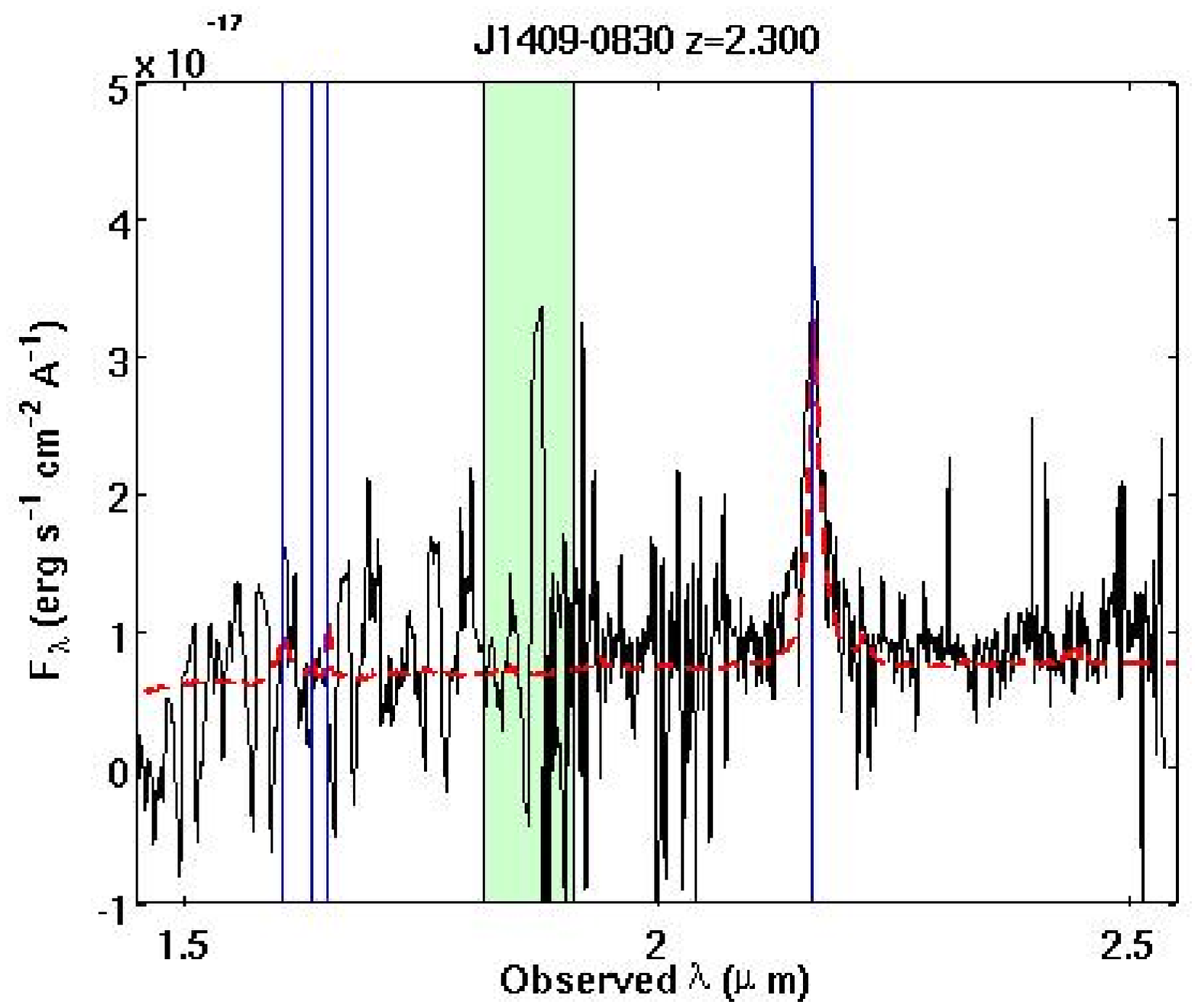} & \includegraphics[scale=0.5,angle=0]{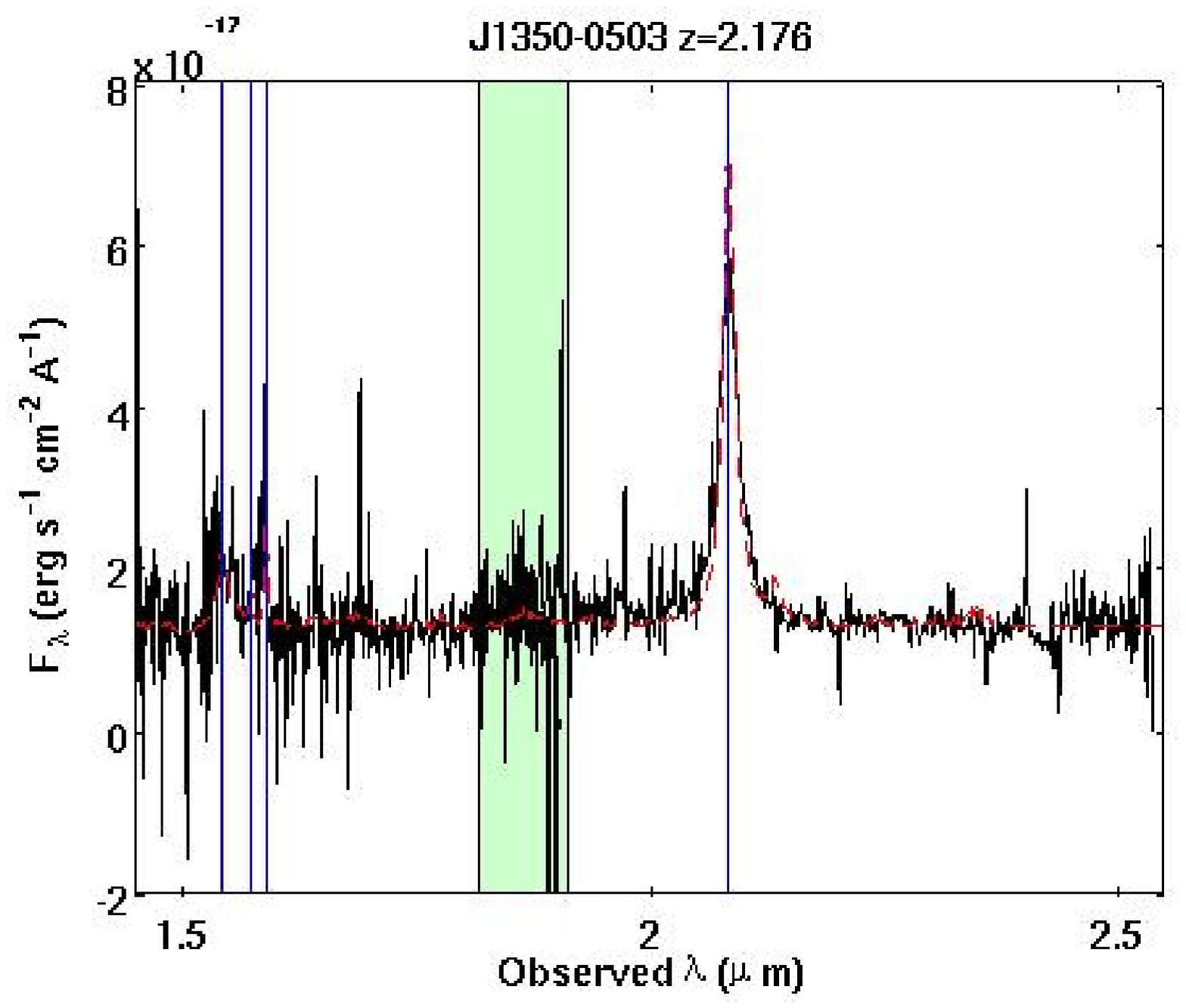} \\
\end{tabular}
%\includegraphics[width=9.5cm,height=7cm,angle=0]{./figs/sky_emission.eps}
%\end{tabular}
\end{minipage}
\caption{NIR spectra of our two confirmed dusty broad-line quasars at z$\sim$2 from VHS+\textit{WISE}. The region
of low atmospheric transmission between the $H$-- and $K$-bands is
shown as the shaded rectangular area. Vertical lines mark the expected positions of emission lines - in order of increasing wavelength - H$\beta$ (4861\AA\@), [OIII] 4959\AA\@, 5007\AA\@ and H$\alpha$ (6563\AA\@).}
\label{fig:spectra}
\end{center}
\end{figure*}

Out of the four reddened quasar candidates observed, two are confirmed to be broad line quasars at z$\sim$2. The spectra can be seen in Figure \ref{fig:spectra} along with the best-fit SEDs derived by fitting to the broadband photometry as in B12. We derive dust extinctions, bolometric luminosities and black-hole masses for these two quasars using the methods outlined in B12. These are summarised in Table \ref{tab:properties}. These properties are fairly typical of our sample of reddened quasars selected from the UKIDSS-LAS (B12). Crucially however, the VHS targets are in the southern hemisphere and therefore suitable for detailed follow-up using ALMA. Optical spectroscopy is needed to confirm the identity of the two sources where no lines were seen in the NIR. These could be highly compact starbursts with buried AGN, or quasars at $z<1.2$, 1.75$<z<$2.00 or $z>2.8$.

\begin{table}
\begin{center}
\caption{Properties of VHS+\textit{WISE} Spectroscopically Confirmed Reddened Quasars}
\label{tab:properties}
\begin{tabular}{lcc}
\hline \hline \\
& VHSJ1350$-$0503 & VHSJ1409$-$0830 \\
\hline
\hline
A$_V$ (mags) & 2.2 & 2.5 \\
log$_{10}$(L$_{\rm{bol}}$/erg s$^{-1}$) & 47.2 & 47.1 \\
log$_{10}$(M$_{\rm{BH}}$/M$_\odot$) & 9.39 & 9.22 \\
Dereddened M$_i$ & $-$28.55 & $-$28.39 \\
$i_{AB}$ from SED Fit & 22.1 & 23.7 \\
\hline 
\end{tabular}
\end{center}
\end{table}

%\begin{table*}
%\begin{center}
%\caption{Properties of New VHS+WISE Reddened Quasars}
%\label{tab:sample}
%\begin{tabular}{lcccc}
%\hline \hline
%Name & A$_V$ & log$_{10}$(L$_{\rm{bol}}$/erg s$^{-1}$) & log$_{10}$(M$_{\rm{BH}}$/M$_\odot$) & lEdd \\
%\hline
%\hline
%VHSATLJ1350-0503 &  &  & & \\
%VHSATLJ1409-0830 &  &  & & \\
%\end{tabular}
%\end{center}
%\end{table*}

\section{MID-INFRARED SEDs OF REDDENED BROAD-LINE QUASARS}

\begin{figure}
\begin{center}
\includegraphics[scale=0.5]{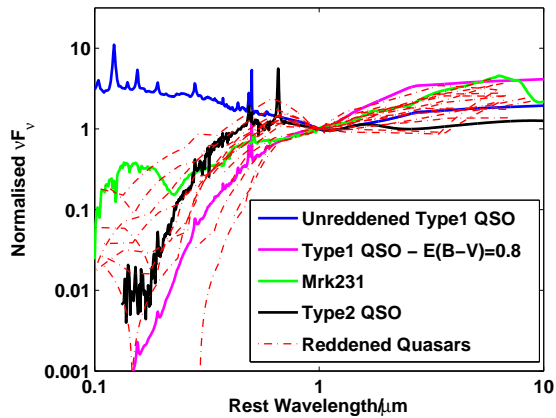}
\caption{Normalised SEDs in terms of $\nu F_\nu$, of all our reddened quasars corrected to the rest-frame and compared to standard SED templates from \citet{Polletta:07}. All SEDs have been normalised to a flux density of 1 at 1$\mu$m.}
\label{fig:sed1}
\end{center}
\end{figure}

We now consider the mid infrared SEDs of all our spectroscopically confirmed reddened quasars including those selected from the UKIDSS-LAS in B12. In Figure \ref{fig:sed1}, we show the SEDs of all 14 spectroscopically confirmed reddened quasars in our sample, and compare them to standard templates from \citet{Polletta:07} including a reddened version of the \citet{Polletta:07} QSO1 template assuming $E(B-V)=0.8$ and an SMC-like extinction curve. The SEDs are all normalised to a value of 1 at 1$\mu$m, the well known inflection point in the quasar SED \citep{Elvis:94}. The full photometric catalogue for all 14 quasars is shown in Table \ref{tab:photometry}. 

We fit two power-laws of the form $\nu F_{\nu} \propto \nu^{\beta}$ to the rest-frame optical SED at 0.3--1$\mu$m as well as the rest-frame infrared SED at 1--3$\mu$m. The QSO1 template has $\beta_{\rm{opt}}=0.68$ and $\beta_{\rm{NIR}}=-$0.55. Mrk231 has optical and infrared power-law exponents of $\beta_{\rm{opt}}=-0.91$ and $\beta_{\rm{NIR}}=-$0.77, while the reddened QSO1 template has $\beta_{\rm{opt}}=-2.27$ and $\beta_{\rm{NIR}}=-$1.26. The rest-frame optical power law exponents in our sample range from $-$6 to $-$0.4 with a median of $\beta_{\rm{opt}}=-$1.47, considerably steeper than the standard QSO1 template. The different values of $\beta_{opt}$ reflect the different dust extinction values in our sample of reddened quasars. The median value of $\beta_{\rm{NIR}}$ in our sample is $-$0.54, consistent with the standard QSO1 template. The infrared slope is therefore not particularly sensitive to dust extinction. Some quasars have steeper infrared slopes consistent with the Mrk231 and reddened QSO1 templates. 
%We note also that the median infrared SED is steeper than the Type 2 QSO template, which has a value of $\beta_{\rm{NIR}}$ close to 0.

The well-known minimum in the quasar SED at $\sim$1$\mu$m \citep{Elvis:94} is clearly seen in the rest-frame SEDs, and these observations demonstrate the importance of this feature on the rest-frame NIR colours. At redshifts of $\sim$0.5 where the 2MASS survey has found significant numbers of red quasars \citep{Glikman:07}, this minimum falls in the $J$-band therefore making the $(J-K)$ colours of the quasars red irrespective of the amount of dust. 

\begin{figure}
\begin{center}
\includegraphics[scale=0.5]{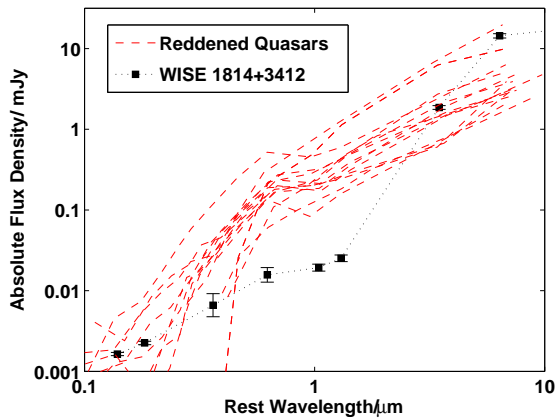}
\caption{Rest-frame SEDs of our reddened quasars compared to the newly discovered hyperluminous infra-red galaxy \textit{WISE} 1814+3412 ($z=2.452$) from \citet{Eisenhardt:12}. Our reddened quasars have comparable W3 fluxes to this HyLIRG at similar redshifts, with two of our most highly reddened sources also showing comparable W4 flux densities.}
\label{fig:sed2}
\end{center}
\end{figure}

In Figure \ref{fig:sed2}, we also compare the \textit{WISE} flux densities of our reddened quasars to the recently discovered Hyperluminous Infrared Galaxy (HyLIRG), \textit{WISE}1814+3412 \citep{Eisenhardt:12} at a similar redshift ($z=2.452$). Our quasars are considerably bluer in terms of their (W2-W3/W4) colours but the median 12$\mu$m flux density is 1.46mJy comparable to 1.86mJy in the case of \textit{WISE}1814+3412. Two of our most highly reddened quasars ULASJ1234+0907 and ULASJ1539+0557 (B12) at $z=2.503$ and $z=2.658$, also have observed 22$\mu$m flux densities of $>$10mJy, comparable to this very luminous HyLIRG. These reddened Type 1 quasars are therefore among the most mid-infrared luminous sources at $z\sim2$, now being discovered using the \textit{WISE} All Sky Survey. 

\begin{table*}
\begin{center}
\caption{AB Magnitudes in the UKIDSS $YJHK$ and \textit{WISE} 3.4, 4.6, 12 and 22$\mu$m bands for all reddened quasars, along with errors on these magnitudes. The full table is published in the online version of this paper.}
\label{tab:photometry}
\begin{tabular}{lcccccccc}
\hline \hline
Name & $Y$ & $J$ & $H$ & $K$ & 3.4$\mu$m & 4.6$\mu$m & 12$\mu$m & 22$\mu$m \\
\hline
\hline
ULASJ1234+0907 & 26.38$\pm$0.84 & 22.38$\pm$0.29 & 20.37$\pm$0.21 & 18.05$\pm$0.03 & 17.20$\pm$0.03 & 16.28$\pm$0.04 & 14.34$\pm$0.04 & 13.90$\pm$0.17 \\
ULASJ1539+0557 & 21.75$\pm$0.36 & 21.02$\pm$0.20 & 19.18$\pm$0.08 & 17.74$\pm$0.02 & 16.85$\pm$0.03 & 16.23$\pm$0.03 & 14.27$\pm$0.03 & 13.18$\pm$0.06 \\
\hline
\textit{Continued online...} & & & & & & & & \\
\end{tabular}
\end{center}
\end{table*}

\section{Conclusions}

We have presented results from a search for reddened broad-line quasars using data from the new infrared VISTA Hemisphere Survey and \textit{WISE} All Sky Survey. We spectroscopically confirm two new reddened quasars in the VHS at z$\sim$2 over an area of 180 deg$^2$. The quasars have broad lines, large bolometric luminosities and black-hole masses, and dust extinctions of A$_V$=2--2.5 mags. We demonstrate that despite being intrinsically very luminous, they are too reddened by dust to be detected in wide-field optical surveys like SDSS. We also present \textit{WISE} photometry for all our reddened broad-line quasars at z$\sim$2, including those selected using the UKIDSS-LAS (B12), as well as the 0.1--10$\mu$m rest-frame SEDs for the entire sample. The rest optical SED power-law slopes reflect the different levels of dust extinction in our quasars, while the rest infrared power-law slopes are similar to Type 1 QSOs with hot dust emission and starburst QSO hosts like Mrk231. The reddened quasars are shown to have similar 12--22$\mu$m flux densities to the recently discovered hyperluminous infrared galaxy, \textit{WISE} 1814+3412 at $z=2.452$. These reddened quasars are therefore among the most mid infrared luminous sources known at these redshifts. Larger samples employing the colour-selection criteria presented here, can now be assembled, and complementary long-wavelength photometry with \textit{Herschel}, SCUBA-2 and ALMA will shed light on the host galaxy properties of these quasars.

\section*{Acknowledgements}

We thank the referee for useful comments. MB, RGM and PCH acknowledge support from the STFC funded Galaxy Formation and Evolution programme at the Institute of Astronomy. Based on observations obtained as part of the VISTA Hemisphere Survey, ESO Program, 179.A-2010 (PI: McMahon) and observations obtained at the Gemini Observatory: GN-2012A-Q-86 (PI:Banerji).

%\bibliographystyle{./reference/mn2e.bst}
%\bibliography{./reference/aamnem99,./reference/eros_ref}

\bibliography{}

\end{document}